\documentclass [a4paper]{JHEP3}
\title{
\protect\vspace{5mm}
Towards the realistic fermion masses with a single
family in extra dimensions.}
\author{Maxim Libanov and Emin~Nougaev\\
Institute for Nuclear Research of the Russian Academy of
Sciences,\\
60th October Anniversary Prospect 7a, 117312, Moscow, Russia.\\
E-mail: \email{ml@ms2.inr.ac.ru, emin@ms2.inr.ac.ru}
}
\abstract{
In a class of multidimensional models, topology of a thick brane
provides three chiral fermionic families with hierarchical masses and
mixings in the effective four-dimensional theory, while the full model
contains a single vector-like generation. We carry out  numerical
simulations and reproduce all known Standard Model fermion masses and
mixings in one of these models.  }

\begin{document}

\section{Introduction} \label{sec:1}

One of interesting features inherent in the theories with more than
four spacetime dimensions is a possibility to explain the mysterious
pattern of fermion mass hierarchies~\cite{AD,LT,LTF} (see also
review~\cite{Rubakov} and references therein). In the previous series of
works~\cite{LT,LTF,neutrino}, we have constructed a model in which a
single family of fermions, with vector-like couplings to the Standard
Model (SM) gauge groups in 6 dimensions, gives rise to three generations
of  chiral Standard Model  fermions in 4 dimensions. This mechanism is
based on  localization of fermion zero modes on a two-dimensional vortex
with winding number $n$. In particular, in the paper~\cite{LT} the case of
global vortex (i.e. group of symmetry of the vortex $U_g(1)$ is global)
with winding number $n=3$ was considered. In this setup three localized
fermionic zero modes appear due to specific Yukawa coupling to the
vortex scalar $\Phi $. Coupling of fermions to the SM Higgs doublet $H$
results in four-dimensional effective fermion masses.  Inter-generation
mixings occur due to explicit breaking of $U_g(1)$ symmetry.  The latter
point is the main drawback of this model since it does not allow to
gauge $U_g(1)$ symmetry. In the paper~\cite{LTF} this problem was
overcome.  The price for unbroken $U_g(1)$, however, is the necessity to
invoke higher dimension operators  in the scalar-fermion
interactions\footnote{The requirement of renormalizability of the theory
does not make much sense anyway in the six-dimensional models, since even
usual Yukawa scalar-fermion-fermion coupling is non-renormalizable.}.
More precisely, to obtain $n$ fermionic generations we consider the vortex
with winding number 1 but fermions  coupled to the vortex scalar
raised to the $n$th power. In both cases (global and gauged $U_g(1)$
symmetry) the hierarchical pattern of the fermion masses occurs due to
different profiles of the fermionic wave functions in the transverse extra
dimensions.  Namely, in Refs.~\cite{LT,LTF} a crude
dimensional analysis has been presented which shows that hierarchical mass
pattern has the form
\begin{equation}
1:g^2:g^4,
\label{OldHierarchy}
\end{equation}
where $g$ is small Yukawa coupling, which yields the localization of
fermionic modes. Also, in Refs.~\cite{LT,LTF} it has been pointed out that
any exact prediction of the fermion masses (as well as a check of the
Eq.(\ref{OldHierarchy})) can be obtained only numerically.

In this paper we return to the models of Refs.~\cite{LT,LTF} and pay
special attention to a check of the Eq.(\ref{OldHierarchy}). We will see
that this equation literally does not hold. Namely,  in
the case of gauged vortex the hierarchical mass pattern takes the form
\[
1:g:g^2.
\]
At the same time, in the case of global vortex unnatural hierarchical
pattern appears.  We will also evaluate fermion masses numerically.  For
reasonable values of parameters in the case of the gauged vortex our
results reproduce SM fermionic mass pattern.  Furthermore, we will obtain
Cabbibo-Kobayashi-Maskawa (CKM) matrix for the same set of parameters, and
see that six-dimensional theory reproduces the mixing matrix quite well.

\section{The Setup} \label{sec:2}

In what follows we restrict ourselves  to  the
model of Ref.~\cite{LTF}. In this section we give a brief description of
this construction.  Our notations coincide with those used in
Refs.~\cite{LT,LTF}. In particular, six-dimensional coordinates $x_A$
are labeled by capital Latin indices $A,B=0,\dots,5$. Four--dimensional
coordinates $x_\mu $ are labeled by Greek indices $\mu,\nu =0,\dots,3$.
The Minkowski metric is $g_{AB}=\mbox{diag}(+,-,\dots,-)$. We use also
chiral representation for six-dimensional Dirac $\Gamma $-matrices (see
Ref.~\cite{LT}).  Besides this we introduce polar coordinates $r$, and
$\theta$ in the $x_4$, $x_5$ plane.

The matter field content of the six-dimensional theory is summarized in
Table \ref{table:fields}. There are three scalar fields. One of these
$\Phi $, together with the $U_g(1)$ gauge field, forms a vortex, while two
other scalars, $X$ and $H$, develop profiles localized on the vortex.  The
potential term of the Lagrangian which gives rise to the non-trivial
profiles for the scalar fields has the following form
\begin{equation}
V_s=
{\lambda\over 2}\left( |\Phi|^2-v^2\right)^2+
{\kappa\over 2} \left( |H|^2-\mu^2\right)^2+h^2|H|^2|\Phi|^2+
{\rho\over 2} \left( |X|^2-v_1^2\right)^2+\eta^2|X|^2|\Phi|^2.
\label{ScalarPotential}
\end{equation}

\TABLE[ht]{
\begin{tabular}{|rc|c|c|c|c|c|}
\hline
\multicolumn{2}{|c|}{fields}
& profiles&\multicolumn{2}{|c|}{charges}&
\multicolumn{2}{|c|}{representations}\\
\cline{4-7}
&&&$U_g(1)$&$U_Y(1)$&$SU_W(2)$&$SU_C(3)$\\
\hline
scalar&$\Phi$&$F(r){\rm e}^{i\theta}$&+1&0&{\bf 1}&{\bf 1}\\
&&$F(0)=0$, $F(\infty)=v$&&&&\\
\hline
scalar&$X$&$X(r)$&+1&0&{\bf 1}&{\bf 1}\\
&&$X(0)=v_X$, $X(\infty)=0$&&&&\\
\hline
scalar&$H$&$H(r)$&$-1$&$+1/2$&{\bf 2}&{\bf 1}\\
&&$H(0)=v_H$, $H(\infty)=0$&&&&\\
\hline
fermion&$Q$&3 L zero modes&axial $+3/2$&$+1/6$&{\bf 2}&{\bf 3}\\
\hline
fermion&$U$&3 R zero modes&axial $-3/2$&$+2/3$&{\bf 1}&{\bf 3}\\
\hline
fermion&$D$&3 R zero modes&axial $-3/2$&$-1/3$&{\bf 1}&{\bf 3}\\
\hline
fermion&$L$&3 L zero modes&axial $+3/2$&$-1/2$&{\bf 2}&{\bf 1}\\
\hline
fermion&$E$&3 R zero modes&axial $-3/2$&$-1$&{\bf 1}&{\bf 1}\\
\hline
\end{tabular}
\caption{Scalars and fermions with their gauge quantum numbers~\cite{LTF}
.
For convenience, we describe here also the profiles of the classical
scalar fields and fermionic wave functions in extra dimensions.}
\label{table:fields}
}

There is also one fermionic generation which consists of five
six-dimensional fermions $Q$, $D$, $U$, $L$ and $E$. These fermions have
vector-like coupling to SM gauge bosons, (a subtle issue of localization
of the latters we do not discuss here), and axial couplings with the
vortex background\footnote{For the scalar--fermion interactions, we take
the most general operators of the lowest order, consistent with  gauge
invariance. On the other hand, in the scalar self--interaction
(\ref{ScalarPotential}) there could be terms of the form $\Phi X^*$ or
$\Phi X^*|H|^2$ which we do not include. These terms, as well as terms of
the form $\Phi ^2X\bar Q (1-\Gamma _7)Q$ in the (\ref{VPhiF}), can be
forbidden, for example, by the imposition of $Z_4$ discrete symmetry:
$\Phi \to i\Phi $, $Q\to \exp(i\Gamma _7\frac{3\pi }{4})Q$, $X\to X$. This
symmetry is explisitly broken by the interaction (\ref{V2}), (\ref{V3}),
and, therefore the undesirable terms will be generated by the radiative
corrections. However, the value of these corrections is of order
$\frac{Y^2}{(4\pi )^3} \epsilon \Lambda ^4$ (for the operator $|H|^2\Phi
X^*$, for instance; $\Lambda $ is the cut off), and for our choice of the
parameters (see Sec. \ref{sec:5}) is of order $10^{-6}\left(\frac{\Lambda
}{M} \right)^4 \lambda $, where $M$ is mass scale, and $\lambda =1/M^2$.
So, if we choose the cut off to be of order $\Lambda \sim 10 M$, then
this corrections can be neglected.}
\begin{equation}
\begin{array}{c}
\displaystyle V_1=g_q\Phi^3\bar Q{1-\Gamma_7\over 2} Q+
g_u\Phi^{*3}\bar U{1-\Gamma_7\over2} U + g_d\Phi^{*3}\bar
D{1-\Gamma_7\over 2} D+\\
\displaystyle g_l\Phi^3\bar L{1-\Gamma_7\over 2} L+ g_e\Phi^{*3}\bar
E{1-\Gamma_7\over2} E+{\rm h.c.}
\end{array}
\label{VPhiF}
\end{equation}
\nopagebreak
($\Gamma _7$ is a six-dimensional analog of four-dimensional matrix
$\gamma _5$).

Note that the scalar background $\Phi^3$, where the field $\Phi$ has
the winding number one, $\Phi=F(r){\rm e}^{i\theta}$, has exactly the same
topological properties as the background $\Phi_1$ of the vortex with
winding number three, $\Phi_1=F_1(r){\rm e}^{3i\theta}$. As a result, the
interactions Eq.~(\ref{VPhiF}) give rise to {\em three} left--handed
(right--handed) zero modes
for each of the fermions $Q$, $L$ ($U$, $D$, $E$). All of these modes are
localized in the core of the vortex\footnote{These modes were discussed in
detail in Ref.~\cite{LT}.}, and could be identified with the three
generations of the chiral SM fermions. The last point, however, is not
quite fair since these {\it zero} modes are {\it massless} from the
four-dimensional point of view. To give small (compared to the vortex
scale $\sqrt{\lambda }v$) masses to these modes we introduce the following
interactions of the fermions with the Higgs doublet $H$ and singlet $X$
\begin{equation}
V_2= Y_dHX\bar Q\frac{1-\Gamma_7}{2}D+ Y_u\tilde HX^*\bar
Q\frac{1-\Gamma_7}{2}U+ Y_lHX\bar L\frac{1-\Gamma_7}{2}E+ {\rm h.c.},
\label{V2}
\end{equation}
where $\tilde H_\alpha =\epsilon_{\alpha \beta }H^*_\beta
$, $\alpha $, $\beta $ are $SU_W(2)$ indices. Finally, there is one
more gauge invariant in the scalar--fermions interaction
\begin{equation}
V_3= Y_d\epsilon_dH\Phi\bar Q\frac{1-\Gamma_7}{2}D+
Y_u\epsilon_u\tilde H\Phi^*\bar Q\frac{1-\Gamma_7}{2}U+
Y_l\epsilon_lH\Phi\bar L\frac{1-\Gamma_7}{2}E+ {\rm h.c.},
\label{V3}
\end{equation}
(we denote Yukawa coupling constants in Eq.~(\ref{V2}) as
$Y_{u,d,l}$, and in Eq.~(\ref{V3}) as $Y_u\epsilon_u, \dots$, for
convenience). This term yields inter-generation mixings.

It is worth noting that to obtain small fermion masses and mixing, it is
not necessary to take $Y$ or $\epsilon $ small. Small masses and mixings
originate from small overlaps of the fermionic wave functions and the
scalar profiles.

Now let us consider this model in more detail. First of all we will
present stable non-trivial configuration for the scalars and
$U_g(1)$ gauge fields. Then we will find fermion zero modes in this
background.  Finally, we will calculate the fermion masses and the CKM
matrix.

\section{The vortex background}\label{sec:3}

As described in the previous section, in our model there are
three scalar fields and one $U_g(1)$ gauge field which form a non-trivial
vortex background. In this section we present a stable solution of the
static field equations which meets all the requirements.

The set of the static equations to the background has the
following form (prime denotes the derivative with respect to $r$; $e$ is
$U_g(1)$ gauge charge),
\begin{eqnarray}
&&F''+{1\over r}F'-{1\over r^2}F(1-A)^2
-\lambda v^2 F \left( F^2-1\right)-h^2FH^2-\eta^2FX^2 =0,\nonumber\\
&&A''-{1\over r}A'+2e^2v^2F^2(1-A)-2e^2H^2A-2e^2X^2A=0,\nonumber\\
&&H''+{1\over r}H'-\frac{1}{r^2}HA^2-h^2 H v^2F^2-\kappa H (H^2-\mu^2)
=0,\nonumber \\
&&X''+{1\over r}X'-\frac{1}{r^2}XA^2-\eta^2 X v^2F^2-\rho X (X^2-{v_1}^2) =0,
\label{qvheq}
\end{eqnarray}
with the boundary conditions
\begin{eqnarray}
&&F(r)= 0,\; H'(r)= 0,\; X'(r)=0, \;A(r)= 0,\qquad {\rm at}\ \ r=
0,\nonumber\\
&&F(r)\to 1,\; H(r)\to 0, \;X(r)\to 0,\; A(r)\to 1,\qquad {\rm at}\ \
r\to\infty.
\label{boundarycondition}
\end{eqnarray}
Here we use the standard anzatz for Abrikosov-Nielsen-Olesen vortex
($i,j=4,5$)
\[
A_{i}(x)=-\frac{1}{er^2}\epsilon _{ij}x_{j}A(r),\ \ \
\Phi(x)=vF(r)e^{i\theta},
\]
\begin{equation}
H_ \alpha (x) = \delta _{2 \alpha }H(r),\ \ \ X(x)=X(r).
\label{anzatz}
\end{equation}

Let us briefly discuss the choice of the boundary conditions
(\ref{boundarycondition}) and the anzatz (\ref{anzatz}). At infinity the
fields should tend to their vacuum expectation values (VEV) which are
determined from the potential (\ref{ScalarPotential}). In general at
arbitrary set of the coupling constants the potential
(\ref{ScalarPotential}) may admit several minima including those in which
$X$ and (or) $H$ are non-zero. However, there are several regions in the
space of the parameters where the potential (\ref{ScalarPotential}) has
{\it only one} minimum $\Phi =v$, $H=X=0$. In particular, one of the
regions is determined by the conditions $h^4>\lambda \kappa $, $\lambda
\rho >\eta ^4$, $\eta ^2v^2>\rho v_1^2$, and in what follows we will
assume that these conditions are satisfied.

In general, any configuration of the fields $\Phi $, $H$, and $X$ may be
characterized by three topological charges $Q_T=(Q_T^\Phi ,Q_T^H, Q_T^X)$.
The zero VEVs of $X$ and $H$ admit the existence of the topologically
trivial configuration of these fields (\ref{anzatz}) (in spite of the
interaction with winded gauge field), $Q_T=(1,0,0)$. In principle, there
may exist a solution of the field equation with $Q_T=(1,-1,1)$ and with
the same topological properties of the gauge field $A_\mu $. However, as
we will see, the solution with $Q_T=(1,0,0)$ is stable, and, therefore,
the choice of the anzatz (\ref{anzatz}) is apropriate.

As  mentioned in the previous section, scalar-fermion interactions
(\ref{V2}), (\ref{V3}) generate low energy mass terms for fermionic modes.
Thus, to obtain non-zero fermion masses one should have a non-trivial
solution for $H(r)$ and $X(r)$.  On the other hand, the boundary value
problem (\ref{qvheq}), (\ref{boundarycondition}) always adopts trivial
solutions $H=0$ or (and) $X=0$. Fortunately, it has been pointed out in
\cite{Witten}, that in some region of parameters these trivial  solutions
are unstable. Namely, the necessary conditions to have non-trivial stable
solutions are $h^2v^2>\kappa\mu^2$ and $\eta^2 v^2>\rho v_1^2$
~\cite{Witten}.  However, it turns out that these conditions are
insufficient. To clarify the situation, let us study the problem of the
stability in more detail.  The stability of the scalar $\Phi $ and gauge
fields is guaranteed by topology.  To investigate the stability of the
static solution $H(r)$ one  linearizes equations of motion
(\ref{qvheq})  and obtain Schr\"{o}dinger-type problem with
the potential
\begin{equation}
h^2v^2F^2(r)+\frac{1}{r^2}A^2-\kappa(\mu^2-3H^2(r)).
\label{shrodinger}
\end{equation}

At large $r$ the solution of  Eq.(\ref{qvheq}), $H(r)$, has
asymptotic $H\sim\exp(-\sqrt{h^2v^2-\kappa\mu^2}\; r)$. So, at large
enough $h$ the Higgs profile becomes narrow  compared to $F(r)$. This
means, in particular, that there is a region of $r$ where $H(r)$ becomes
negligible while $A(r)$ and $F (r)$ are still
almost equal to zero. Therefore, the main contribution to the potential
(\ref{shrodinger}) in this region comes from $r$-independent term $-\kappa
\mu ^2$ and is negative. Thus, the potential becomes not positively
definite, and there  may exist negative levels which  mean the
instability.  The numerical calculations show that this is the case.

\FIGURE[th]{
{
\begingroup%
  \makeatletter%
  \newcommand{\GNUPLOTspecial}{%
    \@sanitize\catcode`\%=14\relax\special}%
  \setlength{\unitlength}{0.1bp}%
{\GNUPLOTspecial{!
/gnudict 256 dict def
gnudict begin
/Color true def
/Solid true def
/gnulinewidth 5.000 def
/userlinewidth gnulinewidth def
/vshift -33 def
/dl {10 mul} def
/hpt_ 31.5 def
/vpt_ 31.5 def
/hpt hpt_ def
/vpt vpt_ def
/M {moveto} bind def
/L {lineto} bind def
/R {rmoveto} bind def
/V {rlineto} bind def
/vpt2 vpt 2 mul def
/hpt2 hpt 2 mul def
/Lshow { currentpoint stroke M
  0 vshift R show } def
/Rshow { currentpoint stroke M
  dup stringwidth pop neg vshift R show } def
/Cshow { currentpoint stroke M
  dup stringwidth pop -2 div vshift R show } def
/UP { dup vpt_ mul /vpt exch def hpt_ mul /hpt exch def
  /hpt2 hpt 2 mul def /vpt2 vpt 2 mul def } def
/DL { Color {setrgbcolor Solid {pop []} if 0 setdash }
 {pop pop pop Solid {pop []} if 0 setdash} ifelse } def
/BL { stroke gnulinewidth 2 mul setlinewidth } def
/AL { stroke gnulinewidth 2 div setlinewidth } def
/UL { gnulinewidth mul /userlinewidth exch def } def
/PL { stroke userlinewidth setlinewidth } def
/LTb { BL [] 0 0 0 DL } def
/LTa { AL [1 dl 2 dl] 0 setdash 0 0 0 setrgbcolor } def
/LT0 { PL [] 1 0 0 DL } def
/LT1 { PL [4 dl 2 dl] 0 1 0 DL } def
/LT2 { PL [2 dl 3 dl] 0 0 1 DL } def
/LT3 { PL [1 dl 1.5 dl] 1 0 1 DL } def
/LT4 { PL [5 dl 2 dl 1 dl 2 dl] 0 1 1 DL } def
/LT5 { PL [4 dl 3 dl 1 dl 3 dl] 1 1 0 DL } def
/LT6 { PL [2 dl 2 dl 2 dl 4 dl] 0 0 0 DL } def
/LT7 { PL [2 dl 2 dl 2 dl 2 dl 2 dl 4 dl] 1 0.3 0 DL } def
/LT8 { PL [2 dl 2 dl 2 dl 2 dl 2 dl 2 dl 2 dl 4 dl] 0.5 0.5 0.5 DL } def
/Pnt { stroke [] 0 setdash
   gsave 1 setlinecap M 0 0 V stroke grestore } def
/Dia { stroke [] 0 setdash 2 copy vpt add M
  hpt neg vpt neg V hpt vpt neg V
  hpt vpt V hpt neg vpt V closepath stroke
  Pnt } def
/Pls { stroke [] 0 setdash vpt sub M 0 vpt2 V
  currentpoint stroke M
  hpt neg vpt neg R hpt2 0 V stroke
  } def
/Box { stroke [] 0 setdash 2 copy exch hpt sub exch vpt add M
  0 vpt2 neg V hpt2 0 V 0 vpt2 V
  hpt2 neg 0 V closepath stroke
  Pnt } def
/Crs { stroke [] 0 setdash exch hpt sub exch vpt add M
  hpt2 vpt2 neg V currentpoint stroke M
  hpt2 neg 0 R hpt2 vpt2 V stroke } def
/TriU { stroke [] 0 setdash 2 copy vpt 1.12 mul add M
  hpt neg vpt -1.62 mul V
  hpt 2 mul 0 V
  hpt neg vpt 1.62 mul V closepath stroke
  Pnt  } def
/Star { 2 copy Pls Crs } def
/BoxF { stroke [] 0 setdash exch hpt sub exch vpt add M
  0 vpt2 neg V  hpt2 0 V  0 vpt2 V
  hpt2 neg 0 V  closepath fill } def
/TriUF { stroke [] 0 setdash vpt 1.12 mul add M
  hpt neg vpt -1.62 mul V
  hpt 2 mul 0 V
  hpt neg vpt 1.62 mul V closepath fill } def
/TriD { stroke [] 0 setdash 2 copy vpt 1.12 mul sub M
  hpt neg vpt 1.62 mul V
  hpt 2 mul 0 V
  hpt neg vpt -1.62 mul V closepath stroke
  Pnt  } def
/TriDF { stroke [] 0 setdash vpt 1.12 mul sub M
  hpt neg vpt 1.62 mul V
  hpt 2 mul 0 V
  hpt neg vpt -1.62 mul V closepath fill} def
/DiaF { stroke [] 0 setdash vpt add M
  hpt neg vpt neg V hpt vpt neg V
  hpt vpt V hpt neg vpt V closepath fill } def
/Pent { stroke [] 0 setdash 2 copy gsave
  translate 0 hpt M 4 {72 rotate 0 hpt L} repeat
  closepath stroke grestore Pnt } def
/PentF { stroke [] 0 setdash gsave
  translate 0 hpt M 4 {72 rotate 0 hpt L} repeat
  closepath fill grestore } def
/Circle { stroke [] 0 setdash 2 copy
  hpt 0 360 arc stroke Pnt } def
/CircleF { stroke [] 0 setdash hpt 0 360 arc fill } def
/C0 { BL [] 0 setdash 2 copy moveto vpt 90 450  arc } bind def
/C1 { BL [] 0 setdash 2 copy        moveto
       2 copy  vpt 0 90 arc closepath fill
               vpt 0 360 arc closepath } bind def
/C2 { BL [] 0 setdash 2 copy moveto
       2 copy  vpt 90 180 arc closepath fill
               vpt 0 360 arc closepath } bind def
/C3 { BL [] 0 setdash 2 copy moveto
       2 copy  vpt 0 180 arc closepath fill
               vpt 0 360 arc closepath } bind def
/C4 { BL [] 0 setdash 2 copy moveto
       2 copy  vpt 180 270 arc closepath fill
               vpt 0 360 arc closepath } bind def
/C5 { BL [] 0 setdash 2 copy moveto
       2 copy  vpt 0 90 arc
       2 copy moveto
       2 copy  vpt 180 270 arc closepath fill
               vpt 0 360 arc } bind def
/C6 { BL [] 0 setdash 2 copy moveto
      2 copy  vpt 90 270 arc closepath fill
              vpt 0 360 arc closepath } bind def
/C7 { BL [] 0 setdash 2 copy moveto
      2 copy  vpt 0 270 arc closepath fill
              vpt 0 360 arc closepath } bind def
/C8 { BL [] 0 setdash 2 copy moveto
      2 copy vpt 270 360 arc closepath fill
              vpt 0 360 arc closepath } bind def
/C9 { BL [] 0 setdash 2 copy moveto
      2 copy  vpt 270 450 arc closepath fill
              vpt 0 360 arc closepath } bind def
/C10 { BL [] 0 setdash 2 copy 2 copy moveto vpt 270 360 arc closepath fill
       2 copy moveto
       2 copy vpt 90 180 arc closepath fill
               vpt 0 360 arc closepath } bind def
/C11 { BL [] 0 setdash 2 copy moveto
       2 copy  vpt 0 180 arc closepath fill
       2 copy moveto
       2 copy  vpt 270 360 arc closepath fill
               vpt 0 360 arc closepath } bind def
/C12 { BL [] 0 setdash 2 copy moveto
       2 copy  vpt 180 360 arc closepath fill
               vpt 0 360 arc closepath } bind def
/C13 { BL [] 0 setdash  2 copy moveto
       2 copy  vpt 0 90 arc closepath fill
       2 copy moveto
       2 copy  vpt 180 360 arc closepath fill
               vpt 0 360 arc closepath } bind def
/C14 { BL [] 0 setdash 2 copy moveto
       2 copy  vpt 90 360 arc closepath fill
               vpt 0 360 arc } bind def
/C15 { BL [] 0 setdash 2 copy vpt 0 360 arc closepath fill
               vpt 0 360 arc closepath } bind def
/Rec   { newpath 4 2 roll moveto 1 index 0 rlineto 0 exch rlineto
       neg 0 rlineto closepath } bind def
/Square { dup Rec } bind def
/Bsquare { vpt sub exch vpt sub exch vpt2 Square } bind def
/S0 { BL [] 0 setdash 2 copy moveto 0 vpt rlineto BL Bsquare } bind def
/S1 { BL [] 0 setdash 2 copy vpt Square fill Bsquare } bind def
/S2 { BL [] 0 setdash 2 copy exch vpt sub exch vpt Square fill Bsquare } bind def
/S3 { BL [] 0 setdash 2 copy exch vpt sub exch vpt2 vpt Rec fill Bsquare } bind def
/S4 { BL [] 0 setdash 2 copy exch vpt sub exch vpt sub vpt Square fill Bsquare } bind def
/S5 { BL [] 0 setdash 2 copy 2 copy vpt Square fill
       exch vpt sub exch vpt sub vpt Square fill Bsquare } bind def
/S6 { BL [] 0 setdash 2 copy exch vpt sub exch vpt sub vpt vpt2 Rec fill Bsquare } bind def
/S7 { BL [] 0 setdash 2 copy exch vpt sub exch vpt sub vpt vpt2 Rec fill
       2 copy vpt Square fill
       Bsquare } bind def
/S8 { BL [] 0 setdash 2 copy vpt sub vpt Square fill Bsquare } bind def
/S9 { BL [] 0 setdash 2 copy vpt sub vpt vpt2 Rec fill Bsquare } bind def
/S10 { BL [] 0 setdash 2 copy vpt sub vpt Square fill 2 copy exch vpt sub exch vpt Square fill
       Bsquare } bind def
/S11 { BL [] 0 setdash 2 copy vpt sub vpt Square fill 2 copy exch vpt sub exch vpt2 vpt Rec fill
       Bsquare } bind def
/S12 { BL [] 0 setdash 2 copy exch vpt sub exch vpt sub vpt2 vpt Rec fill Bsquare } bind def
/S13 { BL [] 0 setdash 2 copy exch vpt sub exch vpt sub vpt2 vpt Rec fill
       2 copy vpt Square fill Bsquare } bind def
/S14 { BL [] 0 setdash 2 copy exch vpt sub exch vpt sub vpt2 vpt Rec fill
       2 copy exch vpt sub exch vpt Square fill Bsquare } bind def
/S15 { BL [] 0 setdash 2 copy Bsquare fill Bsquare } bind def
/D0 { gsave translate 45 rotate 0 0 S0 stroke grestore } bind def
/D1 { gsave translate 45 rotate 0 0 S1 stroke grestore } bind def
/D2 { gsave translate 45 rotate 0 0 S2 stroke grestore } bind def
/D3 { gsave translate 45 rotate 0 0 S3 stroke grestore } bind def
/D4 { gsave translate 45 rotate 0 0 S4 stroke grestore } bind def
/D5 { gsave translate 45 rotate 0 0 S5 stroke grestore } bind def
/D6 { gsave translate 45 rotate 0 0 S6 stroke grestore } bind def
/D7 { gsave translate 45 rotate 0 0 S7 stroke grestore } bind def
/D8 { gsave translate 45 rotate 0 0 S8 stroke grestore } bind def
/D9 { gsave translate 45 rotate 0 0 S9 stroke grestore } bind def
/D10 { gsave translate 45 rotate 0 0 S10 stroke grestore } bind def
/D11 { gsave translate 45 rotate 0 0 S11 stroke grestore } bind def
/D12 { gsave translate 45 rotate 0 0 S12 stroke grestore } bind def
/D13 { gsave translate 45 rotate 0 0 S13 stroke grestore } bind def
/D14 { gsave translate 45 rotate 0 0 S14 stroke grestore } bind def
/D15 { gsave translate 45 rotate 0 0 S15 stroke grestore } bind def
/DiaE { stroke [] 0 setdash vpt add M
  hpt neg vpt neg V hpt vpt neg V
  hpt vpt V hpt neg vpt V closepath stroke } def
/BoxE { stroke [] 0 setdash exch hpt sub exch vpt add M
  0 vpt2 neg V hpt2 0 V 0 vpt2 V
  hpt2 neg 0 V closepath stroke } def
/TriUE { stroke [] 0 setdash vpt 1.12 mul add M
  hpt neg vpt -1.62 mul V
  hpt 2 mul 0 V
  hpt neg vpt 1.62 mul V closepath stroke } def
/TriDE { stroke [] 0 setdash vpt 1.12 mul sub M
  hpt neg vpt 1.62 mul V
  hpt 2 mul 0 V
  hpt neg vpt -1.62 mul V closepath stroke } def
/PentE { stroke [] 0 setdash gsave
  translate 0 hpt M 4 {72 rotate 0 hpt L} repeat
  closepath stroke grestore } def
/CircE { stroke [] 0 setdash
  hpt 0 360 arc stroke } def
/Opaque { gsave closepath 1 setgray fill grestore 0 setgray closepath } def
/DiaW { stroke [] 0 setdash vpt add M
  hpt neg vpt neg V hpt vpt neg V
  hpt vpt V hpt neg vpt V Opaque stroke } def
/BoxW { stroke [] 0 setdash exch hpt sub exch vpt add M
  0 vpt2 neg V hpt2 0 V 0 vpt2 V
  hpt2 neg 0 V Opaque stroke } def
/TriUW { stroke [] 0 setdash vpt 1.12 mul add M
  hpt neg vpt -1.62 mul V
  hpt 2 mul 0 V
  hpt neg vpt 1.62 mul V Opaque stroke } def
/TriDW { stroke [] 0 setdash vpt 1.12 mul sub M
  hpt neg vpt 1.62 mul V
  hpt 2 mul 0 V
  hpt neg vpt -1.62 mul V Opaque stroke } def
/PentW { stroke [] 0 setdash gsave
  translate 0 hpt M 4 {72 rotate 0 hpt L} repeat
  Opaque stroke grestore } def
/CircW { stroke [] 0 setdash
  hpt 0 360 arc Opaque stroke } def
/BoxFill { gsave Rec 1 setgray fill grestore } def
end
}}%
\begin{picture}(3600,2160)(0,0)%
{\GNUPLOTspecial{"
gnudict begin
gsave
0 0 translate
0.100 0.100 scale
0 setgray
newpath
1.000 UL
LTb
400 200 M
63 0 V
2987 0 R
-63 0 V
400 572 M
63 0 V
2987 0 R
-63 0 V
400 944 M
63 0 V
2987 0 R
-63 0 V
400 1316 M
63 0 V
2987 0 R
-63 0 V
400 1688 M
63 0 V
2987 0 R
-63 0 V
400 2060 M
63 0 V
2987 0 R
-63 0 V
400 200 M
0 63 V
0 1797 R
0 -63 V
1163 200 M
0 63 V
0 1797 R
0 -63 V
1925 200 M
0 63 V
0 1797 R
0 -63 V
2688 200 M
0 63 V
0 1797 R
0 -63 V
3450 200 M
0 63 V
0 1797 R
0 -63 V
1.000 UL
LTb
400 200 M
3050 0 V
0 1860 V
-3050 0 V
400 200 L
1.000 UP
1.000 UL
LTb
3087 1947 M
263 0 V
400 200 M
15 30 V
16 30 V
15 30 V
15 30 V
15 30 V
15 30 V
16 30 V
15 30 V
15 30 V
16 29 V
15 30 V
15 30 V
15 29 V
15 30 V
16 29 V
15 29 V
15 29 V
16 29 V
15 29 V
15 29 V
15 28 V
16 28 V
15 28 V
15 28 V
15 28 V
16 27 V
15 28 V
15 27 V
15 26 V
16 27 V
15 26 V
15 26 V
15 25 V
15 25 V
16 25 V
15 25 V
15 24 V
15 24 V
16 23 V
15 23 V
15 23 V
16 22 V
15 22 V
15 21 V
15 21 V
15 21 V
16 20 V
15 19 V
15 20 V
16 18 V
15 19 V
15 18 V
15 17 V
16 17 V
15 17 V
15 16 V
15 16 V
15 15 V
16 15 V
15 14 V
15 14 V
16 14 V
15 13 V
15 13 V
15 12 V
15 12 V
16 11 V
15 11 V
15 11 V
16 11 V
15 10 V
15 9 V
15 10 V
16 9 V
15 9 V
15 8 V
15 8 V
15 8 V
16 7 V
15 7 V
15 7 V
15 7 V
16 6 V
15 7 V
15 6 V
15 5 V
16 6 V
15 5 V
15 5 V
16 5 V
15 4 V
15 5 V
15 4 V
16 4 V
15 4 V
15 4 V
15 4 V
16 3 V
15 3 V
15 3 V
15 3 V
15 3 V
16 3 V
15 3 V
15 2 V
15 3 V
16 2 V
15 2 V
15 3 V
16 2 V
15 2 V
15 1 V
15 2 V
16 2 V
15 2 V
15 1 V
15 2 V
16 1 V
15 1 V
15 2 V
15 1 V
15 1 V
16 1 V
15 2 V
15 1 V
15 1 V
16 1 V
15 1 V
15 0 V
16 1 V
15 1 V
15 1 V
15 1 V
16 0 V
15 1 V
15 1 V
15 0 V
16 1 V
15 0 V
15 1 V
15 0 V
15 1 V
16 0 V
15 1 V
15 0 V
15 1 V
16 0 V
15 1 V
15 0 V
16 0 V
15 1 V
15 0 V
15 0 V
16 1 V
15 0 V
15 0 V
15 0 V
16 1 V
15 0 V
15 0 V
15 0 V
15 1 V
16 0 V
15 0 V
15 0 V
16 0 V
15 0 V
15 1 V
15 0 V
16 0 V
15 0 V
15 0 V
15 0 V
15 0 V
16 1 V
15 0 V
15 0 V
16 0 V
15 0 V
15 0 V
15 0 V
15 0 V
16 0 V
15 1 V
15 0 V
16 0 V
15 0 V
15 0 V
15 0 V
16 0 V
15 0 V
15 0 V
15 0 V
15 0 V
16 0 V
15 0 V
15 0 V
16 0 V
15 0 V
15 1 V
400 200 Pnt
415 230 Pnt
431 260 Pnt
446 290 Pnt
461 320 Pnt
476 350 Pnt
491 380 Pnt
507 410 Pnt
522 440 Pnt
537 470 Pnt
553 499 Pnt
568 529 Pnt
583 559 Pnt
598 588 Pnt
613 618 Pnt
629 647 Pnt
644 676 Pnt
659 705 Pnt
675 734 Pnt
690 763 Pnt
705 792 Pnt
720 820 Pnt
736 848 Pnt
751 876 Pnt
766 904 Pnt
781 932 Pnt
797 959 Pnt
812 987 Pnt
827 1014 Pnt
842 1040 Pnt
858 1067 Pnt
873 1093 Pnt
888 1119 Pnt
903 1144 Pnt
918 1169 Pnt
934 1194 Pnt
949 1219 Pnt
964 1243 Pnt
979 1267 Pnt
995 1290 Pnt
1010 1313 Pnt
1025 1336 Pnt
1041 1358 Pnt
1056 1380 Pnt
1071 1401 Pnt
1086 1422 Pnt
1101 1443 Pnt
1117 1463 Pnt
1132 1482 Pnt
1147 1502 Pnt
1163 1520 Pnt
1178 1539 Pnt
1193 1557 Pnt
1208 1574 Pnt
1224 1591 Pnt
1239 1608 Pnt
1254 1624 Pnt
1269 1640 Pnt
1284 1655 Pnt
1300 1670 Pnt
1315 1684 Pnt
1330 1698 Pnt
1346 1712 Pnt
1361 1725 Pnt
1376 1738 Pnt
1391 1750 Pnt
1406 1762 Pnt
1422 1773 Pnt
1437 1784 Pnt
1452 1795 Pnt
1468 1806 Pnt
1483 1816 Pnt
1498 1825 Pnt
1513 1835 Pnt
1529 1844 Pnt
1544 1853 Pnt
1559 1861 Pnt
1574 1869 Pnt
1589 1877 Pnt
1605 1884 Pnt
1620 1891 Pnt
1635 1898 Pnt
1650 1905 Pnt
1666 1911 Pnt
1681 1918 Pnt
1696 1924 Pnt
1711 1929 Pnt
1727 1935 Pnt
1742 1940 Pnt
1757 1945 Pnt
1773 1950 Pnt
1788 1954 Pnt
1803 1959 Pnt
1818 1963 Pnt
1834 1967 Pnt
1849 1971 Pnt
1864 1975 Pnt
1879 1979 Pnt
1895 1982 Pnt
1910 1985 Pnt
1925 1988 Pnt
1940 1991 Pnt
1955 1994 Pnt
1971 1997 Pnt
1986 2000 Pnt
2001 2002 Pnt
2016 2005 Pnt
2032 2007 Pnt
2047 2009 Pnt
2062 2012 Pnt
2078 2014 Pnt
2093 2016 Pnt
2108 2017 Pnt
2123 2019 Pnt
2139 2021 Pnt
2154 2023 Pnt
2169 2024 Pnt
2184 2026 Pnt
2200 2027 Pnt
2215 2028 Pnt
2230 2030 Pnt
2245 2031 Pnt
2260 2032 Pnt
2276 2033 Pnt
2291 2035 Pnt
2306 2036 Pnt
2321 2037 Pnt
2337 2038 Pnt
2352 2039 Pnt
2367 2039 Pnt
2383 2040 Pnt
2398 2041 Pnt
2413 2042 Pnt
2428 2043 Pnt
2444 2043 Pnt
2459 2044 Pnt
2474 2045 Pnt
2489 2045 Pnt
2505 2046 Pnt
2520 2046 Pnt
2535 2047 Pnt
2550 2047 Pnt
2565 2048 Pnt
2581 2048 Pnt
2596 2049 Pnt
2611 2049 Pnt
2626 2050 Pnt
2642 2050 Pnt
2657 2051 Pnt
2672 2051 Pnt
2688 2051 Pnt
2703 2052 Pnt
2718 2052 Pnt
2733 2052 Pnt
2749 2053 Pnt
2764 2053 Pnt
2779 2053 Pnt
2794 2053 Pnt
2810 2054 Pnt
2825 2054 Pnt
2840 2054 Pnt
2855 2054 Pnt
2870 2055 Pnt
2886 2055 Pnt
2901 2055 Pnt
2916 2055 Pnt
2932 2055 Pnt
2947 2055 Pnt
2962 2056 Pnt
2977 2056 Pnt
2993 2056 Pnt
3008 2056 Pnt
3023 2056 Pnt
3038 2056 Pnt
3053 2056 Pnt
3069 2057 Pnt
3084 2057 Pnt
3099 2057 Pnt
3115 2057 Pnt
3130 2057 Pnt
3145 2057 Pnt
3160 2057 Pnt
3175 2057 Pnt
3191 2057 Pnt
3206 2058 Pnt
3221 2058 Pnt
3237 2058 Pnt
3252 2058 Pnt
3267 2058 Pnt
3282 2058 Pnt
3298 2058 Pnt
3313 2058 Pnt
3328 2058 Pnt
3343 2058 Pnt
3358 2058 Pnt
3374 2058 Pnt
3389 2058 Pnt
3404 2058 Pnt
3420 2058 Pnt
3435 2058 Pnt
3450 2059 Pnt
3218 1947 Pnt
1.000 UL
LT1
3087 1847 M
263 0 V
400 553 M
15 0 V
16 -1 V
15 0 V
15 0 V
15 -1 V
15 0 V
16 -1 V
15 0 V
15 -1 V
16 -1 V
15 -1 V
15 -1 V
15 -1 V
15 -2 V
16 -1 V
15 -1 V
15 -2 V
16 -1 V
15 -2 V
15 -2 V
15 -1 V
16 -2 V
15 -2 V
15 -2 V
15 -2 V
16 -2 V
15 -3 V
15 -2 V
15 -2 V
16 -2 V
15 -3 V
15 -2 V
15 -3 V
15 -2 V
16 -3 V
15 -2 V
15 -3 V
15 -3 V
16 -2 V
15 -3 V
15 -3 V
16 -3 V
15 -2 V
15 -3 V
15 -3 V
15 -3 V
16 -3 V
15 -2 V
15 -3 V
16 -3 V
15 -3 V
15 -3 V
15 -2 V
16 -3 V
15 -3 V
15 -3 V
15 -3 V
15 -3 V
16 -2 V
15 -3 V
15 -3 V
16 -3 V
15 -2 V
15 -3 V
15 -3 V
15 -2 V
16 -3 V
15 -2 V
15 -3 V
16 -3 V
15 -2 V
15 -3 V
15 -2 V
16 -3 V
15 -2 V
15 -2 V
15 -3 V
15 -2 V
16 -2 V
15 -3 V
15 -2 V
15 -2 V
16 -3 V
15 -2 V
15 -2 V
15 -2 V
16 -2 V
15 -2 V
15 -2 V
16 -2 V
15 -2 V
15 -2 V
15 -2 V
16 -2 V
15 -2 V
15 -2 V
15 -2 V
16 -2 V
15 -1 V
15 -2 V
15 -2 V
15 -2 V
16 -1 V
15 -2 V
15 -2 V
15 -1 V
16 -2 V
15 -1 V
15 -2 V
16 -1 V
15 -2 V
15 -2 V
15 -1 V
16 -1 V
15 -2 V
15 -1 V
15 -2 V
16 -1 V
15 -1 V
15 -2 V
15 -1 V
15 -1 V
16 -2 V
15 -1 V
15 -1 V
15 -1 V
16 -2 V
15 -1 V
15 -1 V
16 -1 V
15 -1 V
15 -1 V
15 -1 V
16 -2 V
15 -1 V
15 -1 V
15 -1 V
16 -1 V
15 -1 V
15 -1 V
15 -1 V
15 -1 V
16 -1 V
15 -1 V
15 -1 V
15 -1 V
16 -1 V
15 0 V
15 -1 V
16 -1 V
15 -1 V
15 -1 V
15 -1 V
16 -1 V
15 -1 V
15 0 V
15 -1 V
16 -1 V
15 -1 V
15 -1 V
15 0 V
15 -1 V
16 -1 V
15 -1 V
15 0 V
16 -1 V
15 -1 V
15 0 V
15 -1 V
16 -1 V
15 0 V
15 -1 V
15 -1 V
15 0 V
16 -1 V
15 -1 V
15 0 V
16 -1 V
15 -1 V
15 0 V
15 -1 V
15 0 V
16 -1 V
15 0 V
15 -1 V
16 -1 V
15 0 V
15 -1 V
15 0 V
16 -1 V
15 0 V
15 -1 V
15 0 V
15 -1 V
16 0 V
15 -1 V
15 0 V
16 -1 V
15 0 V
15 -1 V
1.000 UP
1.000 UL
LTa
3087 1747 M
263 0 V
400 200 M
15 0 V
16 0 V
15 0 V
15 1 V
15 0 V
15 1 V
16 0 V
15 1 V
15 1 V
16 1 V
15 1 V
15 1 V
15 1 V
15 1 V
16 2 V
15 1 V
15 2 V
16 1 V
15 2 V
15 2 V
15 2 V
16 2 V
15 2 V
15 2 V
15 2 V
16 2 V
15 3 V
15 2 V
15 3 V
16 2 V
15 3 V
15 3 V
15 3 V
15 3 V
16 3 V
15 3 V
15 3 V
15 3 V
16 3 V
15 3 V
15 4 V
16 3 V
15 4 V
15 3 V
15 4 V
15 4 V
16 3 V
15 4 V
15 4 V
16 4 V
15 4 V
15 4 V
15 4 V
16 4 V
15 4 V
15 4 V
15 4 V
15 4 V
16 4 V
15 5 V
15 4 V
16 4 V
15 5 V
15 4 V
15 5 V
15 4 V
16 5 V
15 4 V
15 5 V
16 4 V
15 5 V
15 4 V
15 5 V
16 5 V
15 4 V
15 5 V
15 5 V
15 5 V
16 4 V
15 5 V
15 5 V
15 5 V
16 5 V
15 5 V
15 4 V
15 5 V
16 5 V
15 5 V
15 5 V
16 5 V
15 5 V
15 5 V
15 5 V
16 5 V
15 5 V
15 5 V
15 4 V
16 5 V
15 5 V
15 5 V
15 5 V
15 5 V
16 5 V
15 5 V
15 5 V
15 5 V
16 5 V
15 5 V
15 5 V
16 5 V
15 5 V
15 5 V
15 5 V
16 5 V
15 5 V
15 5 V
15 5 V
16 5 V
15 5 V
15 5 V
15 5 V
15 5 V
16 5 V
15 5 V
15 5 V
15 5 V
16 5 V
15 5 V
15 5 V
16 5 V
15 5 V
15 5 V
15 5 V
16 5 V
15 5 V
15 5 V
15 5 V
16 4 V
15 5 V
15 5 V
15 5 V
15 5 V
16 5 V
15 5 V
15 5 V
15 4 V
16 5 V
15 5 V
15 5 V
16 5 V
15 5 V
15 4 V
15 5 V
16 5 V
15 5 V
15 4 V
15 5 V
16 5 V
15 5 V
15 4 V
15 5 V
15 5 V
16 5 V
15 4 V
15 5 V
16 5 V
15 4 V
15 5 V
15 4 V
16 5 V
15 5 V
15 4 V
15 5 V
15 4 V
16 5 V
15 5 V
15 4 V
16 5 V
15 4 V
15 5 V
15 4 V
15 5 V
16 4 V
15 5 V
15 4 V
16 5 V
15 4 V
15 4 V
15 5 V
16 4 V
15 5 V
15 4 V
15 4 V
15 5 V
16 4 V
15 4 V
15 5 V
16 4 V
15 4 V
15 5 V
400 200 Pnt
415 200 Pnt
431 200 Pnt
446 200 Pnt
461 201 Pnt
476 201 Pnt
491 202 Pnt
507 202 Pnt
522 203 Pnt
537 204 Pnt
553 205 Pnt
568 206 Pnt
583 207 Pnt
598 208 Pnt
613 209 Pnt
629 211 Pnt
644 212 Pnt
659 214 Pnt
675 215 Pnt
690 217 Pnt
705 219 Pnt
720 221 Pnt
736 223 Pnt
751 225 Pnt
766 227 Pnt
781 229 Pnt
797 231 Pnt
812 234 Pnt
827 236 Pnt
842 239 Pnt
858 241 Pnt
873 244 Pnt
888 247 Pnt
903 250 Pnt
918 253 Pnt
934 256 Pnt
949 259 Pnt
964 262 Pnt
979 265 Pnt
995 268 Pnt
1010 271 Pnt
1025 275 Pnt
1041 278 Pnt
1056 282 Pnt
1071 285 Pnt
1086 289 Pnt
1101 293 Pnt
1117 296 Pnt
1132 300 Pnt
1147 304 Pnt
1163 308 Pnt
1178 312 Pnt
1193 316 Pnt
1208 320 Pnt
1224 324 Pnt
1239 328 Pnt
1254 332 Pnt
1269 336 Pnt
1284 340 Pnt
1300 344 Pnt
1315 349 Pnt
1330 353 Pnt
1346 357 Pnt
1361 362 Pnt
1376 366 Pnt
1391 371 Pnt
1406 375 Pnt
1422 380 Pnt
1437 384 Pnt
1452 389 Pnt
1468 393 Pnt
1483 398 Pnt
1498 402 Pnt
1513 407 Pnt
1529 412 Pnt
1544 416 Pnt
1559 421 Pnt
1574 426 Pnt
1589 431 Pnt
1605 435 Pnt
1620 440 Pnt
1635 445 Pnt
1650 450 Pnt
1666 455 Pnt
1681 460 Pnt
1696 464 Pnt
1711 469 Pnt
1727 474 Pnt
1742 479 Pnt
1757 484 Pnt
1773 489 Pnt
1788 494 Pnt
1803 499 Pnt
1818 504 Pnt
1834 509 Pnt
1849 514 Pnt
1864 519 Pnt
1879 523 Pnt
1895 528 Pnt
1910 533 Pnt
1925 538 Pnt
1940 543 Pnt
1955 548 Pnt
1971 553 Pnt
1986 558 Pnt
2001 563 Pnt
2016 568 Pnt
2032 573 Pnt
2047 578 Pnt
2062 583 Pnt
2078 588 Pnt
2093 593 Pnt
2108 598 Pnt
2123 603 Pnt
2139 608 Pnt
2154 613 Pnt
2169 618 Pnt
2184 623 Pnt
2200 628 Pnt
2215 633 Pnt
2230 638 Pnt
2245 643 Pnt
2260 648 Pnt
2276 653 Pnt
2291 658 Pnt
2306 663 Pnt
2321 668 Pnt
2337 673 Pnt
2352 678 Pnt
2367 683 Pnt
2383 688 Pnt
2398 693 Pnt
2413 698 Pnt
2428 703 Pnt
2444 708 Pnt
2459 713 Pnt
2474 718 Pnt
2489 723 Pnt
2505 727 Pnt
2520 732 Pnt
2535 737 Pnt
2550 742 Pnt
2565 747 Pnt
2581 752 Pnt
2596 757 Pnt
2611 762 Pnt
2626 766 Pnt
2642 771 Pnt
2657 776 Pnt
2672 781 Pnt
2688 786 Pnt
2703 791 Pnt
2718 795 Pnt
2733 800 Pnt
2749 805 Pnt
2764 810 Pnt
2779 814 Pnt
2794 819 Pnt
2810 824 Pnt
2825 829 Pnt
2840 833 Pnt
2855 838 Pnt
2870 843 Pnt
2886 848 Pnt
2901 852 Pnt
2916 857 Pnt
2932 862 Pnt
2947 866 Pnt
2962 871 Pnt
2977 875 Pnt
2993 880 Pnt
3008 885 Pnt
3023 889 Pnt
3038 894 Pnt
3053 898 Pnt
3069 903 Pnt
3084 908 Pnt
3099 912 Pnt
3115 917 Pnt
3130 921 Pnt
3145 926 Pnt
3160 930 Pnt
3175 935 Pnt
3191 939 Pnt
3206 944 Pnt
3221 948 Pnt
3237 953 Pnt
3252 957 Pnt
3267 961 Pnt
3282 966 Pnt
3298 970 Pnt
3313 975 Pnt
3328 979 Pnt
3343 983 Pnt
3358 988 Pnt
3374 992 Pnt
3389 996 Pnt
3404 1001 Pnt
3420 1005 Pnt
3435 1009 Pnt
3450 1014 Pnt
3218 1747 Pnt
1.000 UL
LT6
3087 1647 M
263 0 V
400 2005 M
15 0 V
16 -1 V
15 -1 V
15 -2 V
15 -3 V
15 -3 V
16 -3 V
15 -4 V
15 -4 V
16 -5 V
15 -6 V
15 -6 V
15 -6 V
15 -7 V
16 -8 V
15 -8 V
15 -8 V
16 -10 V
15 -9 V
15 -10 V
15 -10 V
16 -11 V
15 -12 V
15 -12 V
15 -12 V
16 -12 V
15 -14 V
15 -13 V
15 -14 V
16 -14 V
15 -15 V
15 -15 V
15 -15 V
15 -16 V
16 -16 V
15 -16 V
15 -17 V
15 -17 V
16 -17 V
15 -17 V
15 -18 V
16 -18 V
15 -18 V
15 -18 V
15 -18 V
15 -19 V
16 -18 V
15 -19 V
15 -18 V
16 -19 V
15 -19 V
15 -19 V
15 -19 V
16 -19 V
15 -18 V
15 -19 V
15 -19 V
15 -19 V
16 -18 V
15 -19 V
15 -18 V
16 -19 V
15 -18 V
15 -18 V
15 -18 V
15 -18 V
16 -17 V
15 -18 V
15 -17 V
16 -17 V
15 -17 V
15 -17 V
15 -16 V
16 -16 V
15 -16 V
15 -16 V
15 -16 V
15 -15 V
16 -15 V
15 -15 V
15 -15 V
15 -14 V
16 -15 V
15 -14 V
15 -13 V
15 -14 V
16 -13 V
15 -13 V
15 -13 V
16 -13 V
15 -12 V
15 -12 V
15 -12 V
16 -12 V
15 -11 V
15 -11 V
15 -11 V
16 -11 V
15 -11 V
15 -10 V
15 -10 V
15 -10 V
16 -10 V
15 -9 V
15 -9 V
15 -10 V
16 -9 V
15 -8 V
15 -9 V
16 -8 V
15 -8 V
15 -8 V
15 -8 V
16 -8 V
15 -8 V
15 -7 V
15 -7 V
16 -7 V
15 -7 V
15 -7 V
15 -6 V
15 -7 V
16 -6 V
15 -6 V
15 -6 V
15 -6 V
16 -6 V
15 -5 V
15 -6 V
16 -5 V
15 -5 V
15 -6 V
15 -5 V
16 -4 V
15 -5 V
15 -5 V
15 -4 V
16 -5 V
15 -4 V
15 -5 V
15 -4 V
15 -4 V
16 -4 V
15 -4 V
15 -3 V
15 -4 V
16 -4 V
15 -3 V
15 -4 V
16 -3 V
15 -4 V
15 -3 V
15 -3 V
16 -3 V
15 -3 V
15 -3 V
15 -3 V
16 -3 V
15 -2 V
15 -3 V
15 -3 V
15 -2 V
16 -3 V
15 -2 V
15 -3 V
16 -2 V
15 -2 V
15 -3 V
15 -2 V
16 -2 V
15 -2 V
15 -2 V
15 -2 V
15 -2 V
16 -2 V
15 -2 V
15 -2 V
16 -1 V
15 -2 V
15 -2 V
15 -1 V
15 -2 V
16 -2 V
15 -1 V
15 -2 V
16 -1 V
15 -2 V
15 -1 V
15 -1 V
16 -2 V
15 -1 V
15 -1 V
15 -2 V
15 -1 V
16 -1 V
15 -1 V
15 -1 V
16 -1 V
15 -2 V
15 -1 V
stroke
grestore
end
showpage
}}%
\put(3037,1647){\makebox(0,0)[r]{\small $H(r)$}}%
\put(3037,1747){\makebox(0,0)[r]{\small $A(r)$}}%
\put(3037,1847){\makebox(0,0)[r]{\small $X(r)$}}%
\put(3037,1947){\makebox(0,0)[r]{\small $F(r)$}}%
\put(3450,100){\makebox(0,0){20}}%
\put(2688,100){\makebox(0,0){15}}%
\put(1925,100){\makebox(0,0){10}}%
\put(1163,100){\makebox(0,0){5}}%
\put(400,100){\makebox(0,0){0}}%
\put(350,2060){\makebox(0,0)[r]{    1}}%
\put(350,1688){\makebox(0,0)[r]{  0.8}}%
\put(350,1316){\makebox(0,0)[r]{  0.6}}%
\put(350,944){\makebox(0,0)[r]{  0.4}}%
\put(350,572){\makebox(0,0)[r]{  0.2}}%
\put(350,200){\makebox(0,0)[r]{    0}}%
\end{picture}%
\endgroup
 
} \leavevmode
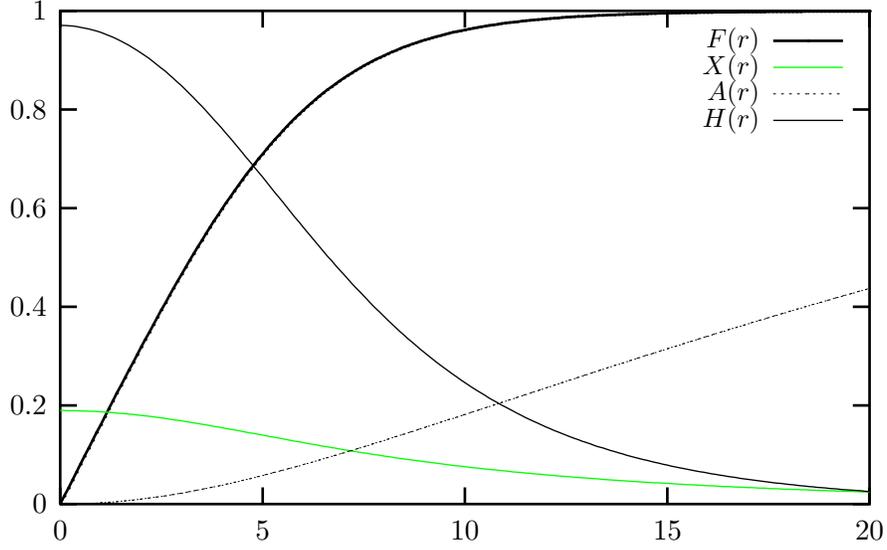
\caption{Profiles of the background at $\lambda=\kappa=\rho=1$, $v=\mu=1$,
$e=0.04$, $h=1.02$, $\eta=0.31$, $v_1=0.3$.
\label{fig1}
} }

To be specific, we have chosen the following set
of the parameters:  $v=\mu=1\cdot
M^2$, $v_1=0.3\cdot M^{2}$, $\lambda=\kappa=\rho=1\cdot M^{-2}$,
$h=1.02\cdot M^{-1}$, $\eta=0.31\cdot M^{-1}$, $e=0.04\cdot M^{-1}$,
where $M$ is a unit of mass which we will define in Section (\ref{sec:5}).
In numerical calculations we assume $M=1$.
To solve the boundary value problem (\ref{qvheq}),
(\ref{boundarycondition}) we have used the relaxation method \cite{Num}.
The solutions for profiles $F(r)$, $H(r)$, $X(r)$ and $A(r)$ in the
vicinity of the origin are presented in Figure \ref{fig1}.
The potential (\ref{shrodinger}) for these solutions is
positive and, therefore, the solutions are stable.

The vortex background in the case of global $U_g(1)$ group can be obtained
from the Eqs.(\ref{qvheq}), (\ref{boundarycondition}) by the setting $e=0$,
$A(r)=0$ ($A(\infty )=0$). The profiles for the scalar fields have the
same shapes as in the gauge case. We do not present the corresponding plot
here.

\section{Mass hierarchy: cases of global and gauged $U_g(1)$}
\label{sec:4}

Now we have all ingredients to find fermionic zero modes in the vortex
background and to obtain hierarchy of the low energy fermionic masses.
To do this let us first briefly remind the analysis
given in Refs.~\cite{LT,LTF}.

\subsection{Global $U_g(1)$}
Let us consider first the case of global $U_g(1)$ group and one
fermionic family, say $Q$ and $U$.
The part of the Lagrangian describing the interaction of
six-dimensional spinor  $Q$ with vortex scalar $\Phi $ is
\begin{equation}
L_Q=i\bar{Q}\Gamma^A\partial_AQ
-\left(g_q\Phi^3\bar Q{1-\Gamma_7\over 2} Q+{\rm h.c.}\right).
\label{ferint}
\end{equation}
This Lagrangian is invariant under  $U_g(1)$ rotations
\[
Q\to{\rm e}^{i{\frac{3\alpha}{2}}\Gamma_7}Q,\,\,\,\Phi \to
{\rm e}^{i\alpha }\Phi.
\]
In the vortex background, one can expand an arbitrary spinor in
eigenfunctions of the corresponding Dirac operator.
There is a set of massive heavy modes (with masses of order $gv^3$) and
zero modes.  The careful analysis given in~\cite{fermionModes,LT,LTF}
shows that in the vortex background with winding number one
there are exactly three localized zero modes  describing four-dimensional
left-handed spinors.  To obtain three right-handed zero modes one should
consider a six-dimensional spinor $U$ which has an axial charge $-3/2$
under $U_g(1)$ with the Lagrangian
\begin{equation}
L_U=i\bar{U}\Gamma^A\partial_AU
-\left(g_u\Phi^3\bar U{1+\Gamma_7\over 2} U+{\rm h.c.}\right).
\label{ferintU}
\end{equation}
These left-handed and right-handed zero modes have the form
\begin{equation}
\begin{array}{c}
\displaystyle
Q_{p}=
\left(
\begin{array}{c}
0 \\
q_{p}(r){\rm e}^{i(3-p)\theta}\\
q_{4-p}(r){\rm e}^{i(1-p)\theta}\\
0 \\
\end{array}
\right)
,\,\,\,
\displaystyle
U_p=
\left(
\begin{array}{c}
u_{4-p}(r){\rm e}^{i(1-p)\theta} \\
0\\
0\\
u_{p}(r){\rm e}^{i(3-p)\theta} \\
\end{array}
\right) \;.
\end{array}
\label{8**}
\end{equation}
Here $p=1,2,3$  enumerates three modes (three fermionic generations). The
radial functions $q_{p}$ are  solutions to the following set of
the equations,
\begin{eqnarray}
&&\displaystyle \left\{
\displaystyle \begin{array}{l}
\displaystyle \partial _rq_{p}-\frac{(3-p)}{r}q_{p}+g_qv^3F^3q_{4-p}=0\\
\\
\displaystyle \partial _rq_{4-p}-\frac{(p-1)}{r}q_{4-p}+g_qv^3F^3q_{p}=0
\end{array}
\right.
\label{ML523}
\end{eqnarray}
with the normalization condition
\begin{equation}
\int\limits_{0}^{\infty
}\!rdr(q_p^2+q_{4-p}^2)=\frac{1}{2\pi}.
\label{norm}
\end{equation}
The functions $u_p(r)$ satisfy the same equations with the
replacement $g_q\to g_u$.

Since the background radial function $F(r)$ is not known in analytical
form, Eqs.(\ref{ML523}) can be solved only numerically. We
will present the numerical results hereinafter, now, however, let us
consider the equations in more details.

We are interested in obtaining low energy fermion masses
which originate from the integration over extra dimensions of the
following expression
\begin{equation}
Y_u\int\!d^2x
\tilde{H}X^*\bar Q\frac{1-\Gamma_7}{2}U +
{\rm h.c.}
\label{massinter}
\end{equation}
Substituting here the zero modes~(\ref{8**}), and integrating over
polar angle $\theta$ one finds
\[
m_{ps}^u = Y_u\int \limits_{}^{}rdrd\theta H(r)X(r)q_{p}(g_q,
r)u_{s}(g_u,r) {\rm e}^{i\theta (r-s)}=
\]
\begin{equation}
\delta _{ps}2\pi Y_u \int\limits_{0}^{\infty
}\!rdr H(r)X(r)q_{p}(g_q, r)u_{p}(g_u,r).
\label{main}
\end{equation}
It is worth noting that due to orthogonality of the zero modes and due to
the fact that $H$ and $X$ do not depend on $\theta $ the integral over
$\theta $ leads to a selection rule, $m_{ps}^u\sim \delta _{ps}$. To
obtain non-trivial  inter-generation mixings one should also consider the
term
\begin{equation}
Y_u\epsilon _u\int\!d^2x
\tilde{H}\Phi ^*\bar Q\frac{1-\Gamma_7}{2}U +
{\rm h.c.}
\label{massintermiss}
\end{equation}
The non-trivial $\theta $-dependence of $\Phi $ (\ref{anzatz})
leads to the off-diagonal elements in the mass matrix
\[
m_{ps}^u = Y_u\epsilon _u\int \limits_{}^{}rdrd\theta H(r)\Phi(r)q_{p}(g_q,
r)u_{s}(g_u,r) {\rm e}^{i\theta (r-s-1)}=
\]
\begin{equation}
\delta _{p,s+1}2\pi Y_u \epsilon _u\int\limits_{0}^{\infty
}\!rdr H(r)\Phi(r)q_{p}(g_q, r)u_{p-1}(g_u,r).
\label{main1}
\end{equation}
In the same way one finds non-zero off-diagonal elements for down-type
quarks $m^d_{ps}\sim\delta _{p+1,s}$. Therefore, the mass matrices have the
following form
\[
m^u=\left(
\begin{array}{ccc}
m_{11}&0&0\\
m_{21}&m_{22}&0\\
0&m_{32}&m_{33}
\end{array}
\right),\ \ \ \
m^d=\left(
\begin{array}{ccc}
m_{11}&m_{12}&0\\
&m_{22}&m_{23}\\
0&0&m_{33}
\end{array}
\right).
\]

To evaluate the integrals (\ref{main}) and (\ref{main1}) over $r$ we take
into account that the functions $H(r)$ and $X(r)$ are localized inside the
vortex. So, the integral (\ref{main}) is saturated near the origin, and
therefore only the behavior of the zero modes at $r\to 0$ is relevant. It
is easy to find from equations (\ref{ML523}) that $q_p\sim r^{3-p}$ at
$r\to 0$, however, it is not so easy to obtain the dependence on $g_q$ of
the zero modes. To obtain the  dependence on small Yukawa couplings $g$ of
the fermion masses we should take into account the normalization condition
(\ref{norm}) and investigate the behavior of the zero modes near the
origin more carefully. To do this let us consider the following simplified
model.

Let us approximate the vortex background by the functions
\[
v^3F(r)^3=\left\{
\begin{array}[]{ll}
r^3,&r<1\\
1,&r\ge 1
\end{array}
\right.
\]
\begin{equation}
H(r)=X(r)=\left\{
\begin{array}[]{ll}
1,&r<1\\
0,&r\ge 1
\end{array}
\right.
\label{aproxhiggs}
\end{equation}
These functions have the correct behavior at the origin and at infinity.
So, this is a reasonable approximation.  The zero mode
equations (\ref{ML523}) with the background Eq.(\ref{aproxhiggs}) have
analytical solution in terms of modified Bessel functions.  However, to
clarify the picture,  we present here only the leading dependence
on $g$ of the solutions  at $r<1$ (which is only relevant for our
purposes), and exact solutions at $r\ge 1$:
\[
q_1(r)=
\left\{
\begin{array}[]{ll}
\tilde{C_1}r^2,&r<1\\
\frac{Cr}{\sqrt{g_q}}e^{-g_qr},&r\ge1 \end{array} \right.  \ \ \ q_2(r)=
\left\{
\begin{array}[]{ll}
\tilde{C_2}r,&r<1\\
\frac{C_mr}{\sqrt{g_q}}e^{-g_qr},&r\ge1
\end{array}
\right.
\]
\begin{equation}
q_3(r)= \left\{
\begin{array}[]{ll}
\tilde{C_3},&r<1\\
\frac{Cr}{\sqrt{g_q}}e^{-g_qr}(1+{1\over{g_qr}}),&r\ge1
\end{array}
\right.
\label{aproxsolution}
\end{equation}

To find the dependence on $g_q$ of the coefficients $C$ and $\tilde{C}$
let us assume that the normalization integral (\ref{norm}) is saturated at
$r\ge 1$. Then substituting the solutions (\ref{aproxsolution}) into the
condition~(\ref{norm}) one has $C\simeq C_m\simeq g_q^{5/2}$. Thus, from
the continuity of the solutions at $r=1$, we find  that
$\tilde{C}_{1,2}\simeq g_q^2$ and $\tilde{C}_3\simeq g_q$. This means in
particular that our assumption is indeed valid:  the normalization
integrals at $r<1$ are of order $g_q^2$ and can be neglected.

Now we can find the mass pattern in this simplified model. Substituting
the solutions (\ref{aproxsolution}) and background (\ref{aproxhiggs}) into
integral (\ref{main}) we have the following dependence on $g_{q,u}$ of the
fermion masses,
\begin{equation}
m_{11}\sim m_{22} \sim (g_q g_u)^2,\ \ \ \ m_{33}\sim (g_q g_u).
\label{globalmasspattern}
\end{equation}
We see that $m_{11}$ and $m_{22}$ are parametrically of the same order,
and therefore there is no hierarchy in the case of global $U_g(1)$. This
result was confirmed by the numerical evaluation of the zero modes and the
integral (\ref{main}).

\subsection{Gauged $U_g(1)$}
Let us concentrate now on the case of the gauged $U_g(1)$. In this case
the equations for the zero modes can be obtained from (\ref{ML523}) by the
replacement
\[
\partial _r\to \partial _r+\frac{3A(r)}{2r}.
\]
Therefore, all dependence of the zero modes on the gauge field  can be
absorbed into a factor
\begin{equation}
(q_p)_{\rm gauge} = \exp{\left(- \int\limits^r {3A(r')
\over 2r'}dr'\right)}\cdot(q_{p})_{\rm global}.
\label{gageoff}
\end{equation}
Approximating  the gauge field as
\begin{equation}
A(r)=
\left\{
\begin{array}[]{ll}
0,&r<1\\
1,&r\ge 1
\end{array}
\right.
\label{aproxforA}
\end{equation}
we find the following zero modes
\[
q_1(r)=
\left\{
\begin{array}[]{ll}
\tilde{C_1}r^2,&r<1\\
\frac{C}{\sqrt{g_qr}}e^{-g_qr},&r\ge1 \end{array} \right.  \ \ \ q_2(r)=
\left\{
\begin{array}[]{ll}
\tilde{C_2}r,&r<1\\
\frac{C_m}{\sqrt{g_qr}}e^{-g_qr},&r\ge1
\end{array}
\right.
\]
\begin{equation}
q_3(r)= \left\{
\begin{array}[]{ll}
\tilde{C_3},&r<1\\
\frac{C}{\sqrt{g_qr}}e^{-g_qr}(1+{1\over{g_qr}}),&r\ge1
\end{array}
\right.
\label{aproxsolutiongauge}
\end{equation}

The situation changes drastically compared to the global
$U_g(1)$ group.  The continuity at $r=1$ results in the following
relations between $\tilde{C_p}$ and $C$, $C_m$:
\begin{equation}
\tilde{C_1}=Cg_q^{-1/2},\qquad \tilde{C_2}=C_mg_q^{-1/2},\qquad
\tilde{C_3}=Cg_q^{-3/2}.
\label{continuitygauge}
\end{equation}
Using the normalization condition (\ref{norm}) one obtains that the
main contribution to the integral (\ref{norm}) for the second mode
comes from $r>1$, and, so, $C_m\sim g_q$.  However, the normalization
integral for the first and third modes is saturated now in the core $r<1$
by the third mode. Thus, $C\sim g_q^{3/2}$ and
\begin{equation}
\tilde{C}_p\sim g_q^{\frac{3-p}{2}}.
\label{cpong}
\end{equation}
With the dependence (\ref{cpong}), one finds the hierarchical
mass pattern in the case of gauge vortex
\begin{equation}
m^u_{pp}\sim Y_u(g_u g_q)^{\frac{3-p}{2}}
\label{masshier}
\end{equation}
which  was announced in the Introduction.

In the same way one estimates mixing terms from the Lagrangian (\ref{V3}),
(\ref{massintermiss})
\begin{equation}
m^u_{ps}\sim\delta _{p,s+1}Y_u\epsilon _u\sqrt{g_u}(g_u
g_q)^{\frac{3-p}{2}},
\label{massmixing}
\end{equation}
and
\begin{equation}
m^u\sim
Y_u\left(
\begin{array}{ccc}
g_u g_q&0&0\\
\epsilon _ug_u\sqrt{g_q}&\sqrt{g_ug_q}&0\\
0&\epsilon _u\sqrt{g_u}&1
\end{array}
\right)
\label{matrixhier}
\end{equation}

\FIGURE[ht]{
   \leavevmode
{
\begingroup%
  \makeatletter%
  \newcommand{\GNUPLOTspecial}{%
    \@sanitize\catcode`\%=14\relax\special}%
  \setlength{\unitlength}{0.1bp}%
{\GNUPLOTspecial{!
/gnudict 256 dict def
gnudict begin
/Color true def
/Solid true def
/gnulinewidth 5.000 def
/userlinewidth gnulinewidth def
/vshift -33 def
/dl {10 mul} def
/hpt_ 31.5 def
/vpt_ 31.5 def
/hpt hpt_ def
/vpt vpt_ def
/M {moveto} bind def
/L {lineto} bind def
/R {rmoveto} bind def
/V {rlineto} bind def
/vpt2 vpt 2 mul def
/hpt2 hpt 2 mul def
/Lshow { currentpoint stroke M
  0 vshift R show } def
/Rshow { currentpoint stroke M
  dup stringwidth pop neg vshift R show } def
/Cshow { currentpoint stroke M
  dup stringwidth pop -2 div vshift R show } def
/UP { dup vpt_ mul /vpt exch def hpt_ mul /hpt exch def
  /hpt2 hpt 2 mul def /vpt2 vpt 2 mul def } def
/DL { Color {setrgbcolor Solid {pop []} if 0 setdash }
 {pop pop pop Solid {pop []} if 0 setdash} ifelse } def
/BL { stroke gnulinewidth 2 mul setlinewidth } def
/AL { stroke gnulinewidth 2 div setlinewidth } def
/UL { gnulinewidth mul /userlinewidth exch def } def
/PL { stroke userlinewidth setlinewidth } def
/LTb { BL [] 0 0 0 DL } def
/LTa { AL [1 dl 2 dl] 0 setdash 0 0 0 setrgbcolor } def
/LT0 { PL [] 1 0 0 DL } def
/LT1 { PL [4 dl 2 dl] 0 1 0 DL } def
/LT2 { PL [2 dl 3 dl] 0 0 1 DL } def
/LT3 { PL [1 dl 1.5 dl] 1 0 1 DL } def
/LT4 { PL [5 dl 2 dl 1 dl 2 dl] 0 1 1 DL } def
/LT5 { PL [4 dl 3 dl 1 dl 3 dl] 1 1 0 DL } def
/LT6 { PL [2 dl 2 dl 2 dl 4 dl] 0 0 0 DL } def
/LT7 { PL [2 dl 2 dl 2 dl 2 dl 2 dl 4 dl] 1 0.3 0 DL } def
/LT8 { PL [2 dl 2 dl 2 dl 2 dl 2 dl 2 dl 2 dl 4 dl] 0.5 0.5 0.5 DL } def
/Pnt { stroke [] 0 setdash
   gsave 1 setlinecap M 0 0 V stroke grestore } def
/Dia { stroke [] 0 setdash 2 copy vpt add M
  hpt neg vpt neg V hpt vpt neg V
  hpt vpt V hpt neg vpt V closepath stroke
  Pnt } def
/Pls { stroke [] 0 setdash vpt sub M 0 vpt2 V
  currentpoint stroke M
  hpt neg vpt neg R hpt2 0 V stroke
  } def
/Box { stroke [] 0 setdash 2 copy exch hpt sub exch vpt add M
  0 vpt2 neg V hpt2 0 V 0 vpt2 V
  hpt2 neg 0 V closepath stroke
  Pnt } def
/Crs { stroke [] 0 setdash exch hpt sub exch vpt add M
  hpt2 vpt2 neg V currentpoint stroke M
  hpt2 neg 0 R hpt2 vpt2 V stroke } def
/TriU { stroke [] 0 setdash 2 copy vpt 1.12 mul add M
  hpt neg vpt -1.62 mul V
  hpt 2 mul 0 V
  hpt neg vpt 1.62 mul V closepath stroke
  Pnt  } def
/Star { 2 copy Pls Crs } def
/BoxF { stroke [] 0 setdash exch hpt sub exch vpt add M
  0 vpt2 neg V  hpt2 0 V  0 vpt2 V
  hpt2 neg 0 V  closepath fill } def
/TriUF { stroke [] 0 setdash vpt 1.12 mul add M
  hpt neg vpt -1.62 mul V
  hpt 2 mul 0 V
  hpt neg vpt 1.62 mul V closepath fill } def
/TriD { stroke [] 0 setdash 2 copy vpt 1.12 mul sub M
  hpt neg vpt 1.62 mul V
  hpt 2 mul 0 V
  hpt neg vpt -1.62 mul V closepath stroke
  Pnt  } def
/TriDF { stroke [] 0 setdash vpt 1.12 mul sub M
  hpt neg vpt 1.62 mul V
  hpt 2 mul 0 V
  hpt neg vpt -1.62 mul V closepath fill} def
/DiaF { stroke [] 0 setdash vpt add M
  hpt neg vpt neg V hpt vpt neg V
  hpt vpt V hpt neg vpt V closepath fill } def
/Pent { stroke [] 0 setdash 2 copy gsave
  translate 0 hpt M 4 {72 rotate 0 hpt L} repeat
  closepath stroke grestore Pnt } def
/PentF { stroke [] 0 setdash gsave
  translate 0 hpt M 4 {72 rotate 0 hpt L} repeat
  closepath fill grestore } def
/Circle { stroke [] 0 setdash 2 copy
  hpt 0 360 arc stroke Pnt } def
/CircleF { stroke [] 0 setdash hpt 0 360 arc fill } def
/C0 { BL [] 0 setdash 2 copy moveto vpt 90 450  arc } bind def
/C1 { BL [] 0 setdash 2 copy        moveto
       2 copy  vpt 0 90 arc closepath fill
               vpt 0 360 arc closepath } bind def
/C2 { BL [] 0 setdash 2 copy moveto
       2 copy  vpt 90 180 arc closepath fill
               vpt 0 360 arc closepath } bind def
/C3 { BL [] 0 setdash 2 copy moveto
       2 copy  vpt 0 180 arc closepath fill
               vpt 0 360 arc closepath } bind def
/C4 { BL [] 0 setdash 2 copy moveto
       2 copy  vpt 180 270 arc closepath fill
               vpt 0 360 arc closepath } bind def
/C5 { BL [] 0 setdash 2 copy moveto
       2 copy  vpt 0 90 arc
       2 copy moveto
       2 copy  vpt 180 270 arc closepath fill
               vpt 0 360 arc } bind def
/C6 { BL [] 0 setdash 2 copy moveto
      2 copy  vpt 90 270 arc closepath fill
              vpt 0 360 arc closepath } bind def
/C7 { BL [] 0 setdash 2 copy moveto
      2 copy  vpt 0 270 arc closepath fill
              vpt 0 360 arc closepath } bind def
/C8 { BL [] 0 setdash 2 copy moveto
      2 copy vpt 270 360 arc closepath fill
              vpt 0 360 arc closepath } bind def
/C9 { BL [] 0 setdash 2 copy moveto
      2 copy  vpt 270 450 arc closepath fill
              vpt 0 360 arc closepath } bind def
/C10 { BL [] 0 setdash 2 copy 2 copy moveto vpt 270 360 arc closepath fill
       2 copy moveto
       2 copy vpt 90 180 arc closepath fill
               vpt 0 360 arc closepath } bind def
/C11 { BL [] 0 setdash 2 copy moveto
       2 copy  vpt 0 180 arc closepath fill
       2 copy moveto
       2 copy  vpt 270 360 arc closepath fill
               vpt 0 360 arc closepath } bind def
/C12 { BL [] 0 setdash 2 copy moveto
       2 copy  vpt 180 360 arc closepath fill
               vpt 0 360 arc closepath } bind def
/C13 { BL [] 0 setdash  2 copy moveto
       2 copy  vpt 0 90 arc closepath fill
       2 copy moveto
       2 copy  vpt 180 360 arc closepath fill
               vpt 0 360 arc closepath } bind def
/C14 { BL [] 0 setdash 2 copy moveto
       2 copy  vpt 90 360 arc closepath fill
               vpt 0 360 arc } bind def
/C15 { BL [] 0 setdash 2 copy vpt 0 360 arc closepath fill
               vpt 0 360 arc closepath } bind def
/Rec   { newpath 4 2 roll moveto 1 index 0 rlineto 0 exch rlineto
       neg 0 rlineto closepath } bind def
/Square { dup Rec } bind def
/Bsquare { vpt sub exch vpt sub exch vpt2 Square } bind def
/S0 { BL [] 0 setdash 2 copy moveto 0 vpt rlineto BL Bsquare } bind def
/S1 { BL [] 0 setdash 2 copy vpt Square fill Bsquare } bind def
/S2 { BL [] 0 setdash 2 copy exch vpt sub exch vpt Square fill Bsquare } bind def
/S3 { BL [] 0 setdash 2 copy exch vpt sub exch vpt2 vpt Rec fill Bsquare } bind def
/S4 { BL [] 0 setdash 2 copy exch vpt sub exch vpt sub vpt Square fill Bsquare } bind def
/S5 { BL [] 0 setdash 2 copy 2 copy vpt Square fill
       exch vpt sub exch vpt sub vpt Square fill Bsquare } bind def
/S6 { BL [] 0 setdash 2 copy exch vpt sub exch vpt sub vpt vpt2 Rec fill Bsquare } bind def
/S7 { BL [] 0 setdash 2 copy exch vpt sub exch vpt sub vpt vpt2 Rec fill
       2 copy vpt Square fill
       Bsquare } bind def
/S8 { BL [] 0 setdash 2 copy vpt sub vpt Square fill Bsquare } bind def
/S9 { BL [] 0 setdash 2 copy vpt sub vpt vpt2 Rec fill Bsquare } bind def
/S10 { BL [] 0 setdash 2 copy vpt sub vpt Square fill 2 copy exch vpt sub exch vpt Square fill
       Bsquare } bind def
/S11 { BL [] 0 setdash 2 copy vpt sub vpt Square fill 2 copy exch vpt sub exch vpt2 vpt Rec fill
       Bsquare } bind def
/S12 { BL [] 0 setdash 2 copy exch vpt sub exch vpt sub vpt2 vpt Rec fill Bsquare } bind def
/S13 { BL [] 0 setdash 2 copy exch vpt sub exch vpt sub vpt2 vpt Rec fill
       2 copy vpt Square fill Bsquare } bind def
/S14 { BL [] 0 setdash 2 copy exch vpt sub exch vpt sub vpt2 vpt Rec fill
       2 copy exch vpt sub exch vpt Square fill Bsquare } bind def
/S15 { BL [] 0 setdash 2 copy Bsquare fill Bsquare } bind def
/D0 { gsave translate 45 rotate 0 0 S0 stroke grestore } bind def
/D1 { gsave translate 45 rotate 0 0 S1 stroke grestore } bind def
/D2 { gsave translate 45 rotate 0 0 S2 stroke grestore } bind def
/D3 { gsave translate 45 rotate 0 0 S3 stroke grestore } bind def
/D4 { gsave translate 45 rotate 0 0 S4 stroke grestore } bind def
/D5 { gsave translate 45 rotate 0 0 S5 stroke grestore } bind def
/D6 { gsave translate 45 rotate 0 0 S6 stroke grestore } bind def
/D7 { gsave translate 45 rotate 0 0 S7 stroke grestore } bind def
/D8 { gsave translate 45 rotate 0 0 S8 stroke grestore } bind def
/D9 { gsave translate 45 rotate 0 0 S9 stroke grestore } bind def
/D10 { gsave translate 45 rotate 0 0 S10 stroke grestore } bind def
/D11 { gsave translate 45 rotate 0 0 S11 stroke grestore } bind def
/D12 { gsave translate 45 rotate 0 0 S12 stroke grestore } bind def
/D13 { gsave translate 45 rotate 0 0 S13 stroke grestore } bind def
/D14 { gsave translate 45 rotate 0 0 S14 stroke grestore } bind def
/D15 { gsave translate 45 rotate 0 0 S15 stroke grestore } bind def
/DiaE { stroke [] 0 setdash vpt add M
  hpt neg vpt neg V hpt vpt neg V
  hpt vpt V hpt neg vpt V closepath stroke } def
/BoxE { stroke [] 0 setdash exch hpt sub exch vpt add M
  0 vpt2 neg V hpt2 0 V 0 vpt2 V
  hpt2 neg 0 V closepath stroke } def
/TriUE { stroke [] 0 setdash vpt 1.12 mul add M
  hpt neg vpt -1.62 mul V
  hpt 2 mul 0 V
  hpt neg vpt 1.62 mul V closepath stroke } def
/TriDE { stroke [] 0 setdash vpt 1.12 mul sub M
  hpt neg vpt 1.62 mul V
  hpt 2 mul 0 V
  hpt neg vpt -1.62 mul V closepath stroke } def
/PentE { stroke [] 0 setdash gsave
  translate 0 hpt M 4 {72 rotate 0 hpt L} repeat
  closepath stroke grestore } def
/CircE { stroke [] 0 setdash
  hpt 0 360 arc stroke } def
/Opaque { gsave closepath 1 setgray fill grestore 0 setgray closepath } def
/DiaW { stroke [] 0 setdash vpt add M
  hpt neg vpt neg V hpt vpt neg V
  hpt vpt V hpt neg vpt V Opaque stroke } def
/BoxW { stroke [] 0 setdash exch hpt sub exch vpt add M
  0 vpt2 neg V hpt2 0 V 0 vpt2 V
  hpt2 neg 0 V Opaque stroke } def
/TriUW { stroke [] 0 setdash vpt 1.12 mul add M
  hpt neg vpt -1.62 mul V
  hpt 2 mul 0 V
  hpt neg vpt 1.62 mul V Opaque stroke } def
/TriDW { stroke [] 0 setdash vpt 1.12 mul sub M
  hpt neg vpt 1.62 mul V
  hpt 2 mul 0 V
  hpt neg vpt -1.62 mul V Opaque stroke } def
/PentW { stroke [] 0 setdash gsave
  translate 0 hpt M 4 {72 rotate 0 hpt L} repeat
  Opaque stroke grestore } def
/CircW { stroke [] 0 setdash
  hpt 0 360 arc Opaque stroke } def
/BoxFill { gsave Rec 1 setgray fill grestore } def
end
}}%
\begin{picture}(3600,2160)(0,0)%
{\GNUPLOTspecial{"
gnudict begin
gsave
0 0 translate
0.100 0.100 scale
0 setgray
newpath
1.000 UL
LTb
150 200 M
0 63 V
0 1797 R
0 -63 V
647 200 M
0 31 V
0 1829 R
0 -31 V
937 200 M
0 31 V
0 1829 R
0 -31 V
1143 200 M
0 31 V
0 1829 R
0 -31 V
1303 200 M
0 31 V
0 1829 R
0 -31 V
1434 200 M
0 31 V
0 1829 R
0 -31 V
1544 200 M
0 31 V
0 1829 R
0 -31 V
1640 200 M
0 31 V
0 1829 R
0 -31 V
1725 200 M
0 31 V
0 1829 R
0 -31 V
1800 200 M
0 63 V
0 1797 R
0 -63 V
2297 200 M
0 31 V
0 1829 R
0 -31 V
2587 200 M
0 31 V
0 1829 R
0 -31 V
2793 200 M
0 31 V
0 1829 R
0 -31 V
2953 200 M
0 31 V
0 1829 R
0 -31 V
3084 200 M
0 31 V
0 1829 R
0 -31 V
3194 200 M
0 31 V
0 1829 R
0 -31 V
3290 200 M
0 31 V
0 1829 R
0 -31 V
3375 200 M
0 31 V
0 1829 R
0 -31 V
3450 200 M
0 63 V
0 1797 R
0 -63 V
1.000 UL
LTb
150 200 M
3300 0 V
0 1860 V
-3300 0 V
150 200 L
1.000 UP
1.000 UL
LT1
3087 1947 M
263 0 V
647 1794 M
496 0 V
291 0 V
206 0 V
160 0 V
68 0 V
63 0 V
160 0 V
647 1794 TriDF
1143 1794 TriDF
1434 1794 TriDF
1640 1794 TriDF
1800 1794 TriDF
1868 1794 TriDF
1931 1794 TriDF
2091 1794 TriDF
3218 1947 TriDF
1.000 UP
1.000 UL
LT2
3087 1847 M
263 0 V
647 1449 M
496 13 V
291 0 V
206 13 V
160 0 V
68 0 V
63 0 V
160 0 V
647 1449 Dia
1143 1462 Dia
1434 1462 Dia
1640 1475 Dia
1800 1475 Dia
1868 1475 Dia
1931 1475 Dia
2091 1475 Dia
3218 1847 Dia
1.000 UP
1.000 UL
LT3
3087 1747 M
263 0 V
647 785 M
496 13 V
291 13 V
206 13 V
160 11 V
68 5 V
63 3 V
160 21 V
647 785 DiaF
1143 798 DiaF
1434 811 DiaF
1640 824 DiaF
1800 835 DiaF
1868 840 DiaF
1931 843 DiaF
2091 864 DiaF
3218 1747 DiaF
1.000 UP
1.000 UL
LT4
3087 1647 M
263 0 V
647 690 M
496 -12 V
291 -13 V
206 -11 V
160 -13 V
68 -5 V
63 -7 V
160 -17 V
647 690 Pent
1143 678 Pent
1434 665 Pent
1640 654 Pent
1800 641 Pent
1868 636 Pent
1931 629 Pent
2091 612 Pent
3218 1647 Pent
1.000 UP
1.000 UL
LT5
3087 1547 M
263 0 V
647 436 M
496 -10 V
291 -9 V
206 -9 V
160 -8 V
68 -4 V
63 -4 V
160 -11 V
647 436 PentF
1143 426 PentF
1434 417 PentF
1640 408 PentF
1800 400 PentF
1868 396 PentF
1931 392 PentF
2091 381 PentF
3218 1547 PentF
stroke
grestore
end
showpage
}}%
\put(3037,1547){\makebox(0,0)[r]{\small $\tilde{m}_{11}$}}%
\put(3037,1647){\makebox(0,0)[r]{\small $\tilde{m}_{21}$}}%
\put(3037,1747){\makebox(0,0)[r]{\small $\tilde{m}_{22}$}}%
\put(3037,1847){\makebox(0,0)[r]{\small $\tilde{m}_{32}$}}%
\put(3037,1947){\makebox(0,0)[r]{\small $\tilde{m}_{33}$}}%
\put(3450,100){\makebox(0,0){0.1}}%
\put(1800,100){\makebox(0,0){0.01}}%
\put(150,100){\makebox(0,0){0.001}}%
\end{picture}%
\endgroup
 
}
\caption{
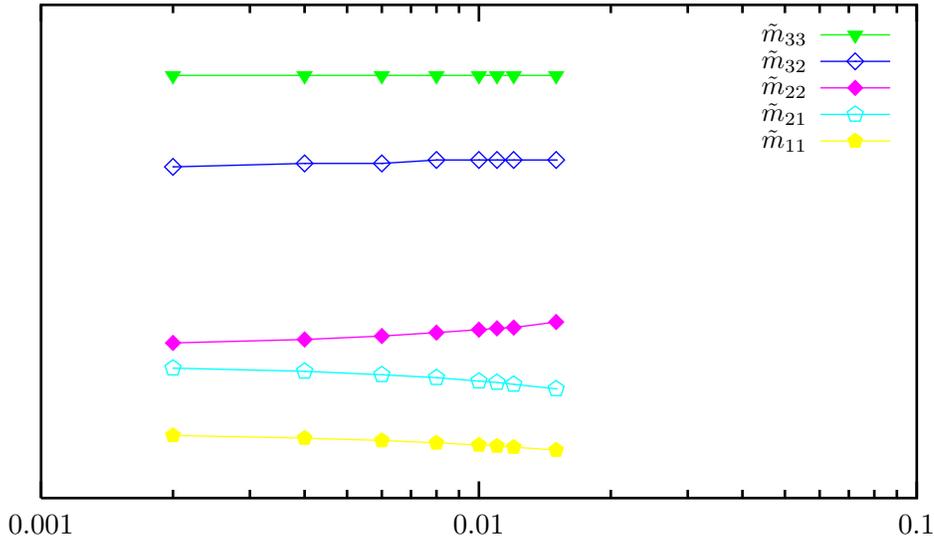
The $g$-dependence of the fermion mass pattern
$\tilde{m}_{ps}=m_{ps}/g^{(6-p-s)/2}$ calculated in the background
shown at Fig.  \ref{fig1}.
\label{fig2}
}
}
Of course, the vortex background obtained numerically
differs from our crude approximation. In particular, one can see
from  Fig. \ref{fig1} that the gauge field $A(r)$ is wider than the
scalar $F (r)$, and even far outside the core is almost equal to zero.
So, our approximation (\ref{aproxforA}) is not quite appropriate.  One may
expect that in the real background (Fig.  \ref{fig1}) the $g$-dependence
of the mass pattern  is more resembling with the dependence in the case of
the global $U_g(1)$ group (\ref{globalmasspattern}) when $A(r)=0$.
However, the numerical calculations show that the mass pattern depends
weakly on the width of the gauge field. We present in  Fig.~\ref{fig2}
five non-zero elements of the mass matrix
(\ref{matrixhier}) divided by the corresponding power of
$g$ in the case $g=g_q=g_u$. We see that $m_{ps}/g^{(6-p-s)/2}$  indeed
depend weakly on $g$ in the wide interesting range.

\section{Results for Standard Model fermions}\label{sec:5}

Now we are ready to find the set of the parameters of the our model ($Y$,
$\epsilon $, $g$) which reproduce the real SM fermionic mass pattern.

To do this, one has to find  eigenvalues of the mass matrix
$m^u_{ps}(Y_u,\epsilon_u ,g_q,g_u)$, to equate the result with known
fermion masses, and to obtain three equations on four unknown variable,
$Y_u$, $\epsilon _u$, $g_q$, $g_u$. However,  even to solve numerically
these equations is not an easy problem.  The point is that to find the
solution one has to obtain the mass matrix $m_{ps}^u$ in some range of the
parameters, which requires in turn the calculation of the zero modes
(solving the Eqs.  (\ref{ML523})) at each value of $g$.  Fortunately,
there is a simpler way. First of all, note that the mass matrix is
proportional to $Y_{u,(d,l)}$, and, therefore, any ratio of the masses does
not depend on $Y_{u,(d,l)}$.  Forming two ratios of the masses one obtains two
equations on three variables $\epsilon _u$, $g_q$, $g_u$.  The second
simplification is to use the established dependence of the mass matrix on
$g$ and $\epsilon $ (\ref{masshier}), (\ref{massmixing}). After this the
obtained equations can be easily solved.  In fact, in the quark sector
the situation is more complicated since besides the mass matrix for
up-quarks,  there are also the mass matrix for down quarks and the
CKM matrix.  Therefore, we have seven equations (four equations for the
mass ratios and three equations for CKM angles) on five parameters ($g_q$,
$g_u$, $g_d$, $\epsilon _u$, $\epsilon _d$). In our simulations we have
used five of seven equations to reproduce correct mass ratios and one
element of the CKM-matrix ($U_{33}^{CKM}$).  As a result we have
obtained
\begin{equation}
g_q = 0.0013\cdot M^{-5},\ \ g_u = 0.0013\cdot M^{-5},\ \ g_d=0.03\cdot
M^{-5}, \ \ \epsilon _u=0.9, \ \ \epsilon _d=0.1.
\label{g-our}
\end{equation}
Having at hand these parameters, one can find two other
independent elements of the CKM-matrix, but, in general, we could not
expect that they  will coincide with the known values.  However, we
have obtained the following CKM-matrix
\begin{equation}
U^{CKM}= \left(
\begin{array}{ccc}
0.9780& 0.2084& 0.0100\\
0.2086& 0.9769& 0.0453\\
0.0004& 0.0464& 0.9990
\end{array}
\right),
\label{ckm}
\end{equation}
and see that this matrix coincides quite well with the measured CKM-matrix
in the Standard Model~\cite{PDG}
\begin{equation}
U_{SM}^{CKM}= \left(
\begin{array}{lll}
0.9742 \div 0.9757& 0.219 \div 0.226& 0.002 \div
0.005\\
0.219 \div 0.225& 0.9734 \div 0.9749& 0.037 \div 0.043\\
0.004 \div 0.014& 0.035 \div 0.043& 0.9990 \div 0.9993
\end{array}
\right).
\label{ckm1}
\end{equation}

In the leptonic sector the correct mass ratios are reproduced at
\[
g_l= g_e= 0.01,\ \  \epsilon_l = 0.75.
\]

To find Yukawa couplings $Y$ we should first of all understand how the
obtained  dimensionful parameters (that is all parameters except
$\epsilon $) correlate with the absolute energy scale. Let us assume that
the first non-zero fermion level is of order 100TeV. It means that
$gv^3\simeq 100$TeV. Substituting here $g=0.001$ and $v=1$ one finds our
units:
\[
1\cdot M=10^5\mbox{TeV}.
\]
With this relation, one can convert all parameters measured in
our units to the absolute mass scale:
\[
v=\mu =10^{10}{\rm TeV}^2,\ \  v_1=3\cdot 10^{9}{\rm TeV}^2
\]
\[
\lambda =\kappa =\rho =10^{-10}{\rm TeV}^{-2},\ \ h=1.02\cdot 10^{-5}
{\rm TeV}^{-1}, \ \ \eta =3.1\cdot 10^{-6}{\rm TeV}^{-1}, \ \
e=4\cdot 10^{-7}{\rm TeV}^{-1}
\]
\[
g_q=g_u=1.3\cdot 10^{-28}{\rm TeV}^{-5},\ \ g_d=3\cdot 10^{-27}{\rm
TeV}^{-5},\ \ g_l=g_e=1\cdot 10^{-27}{\rm TeV}^{-5}.
\]
Now taking into account the actual values of the fermion masses one
finds the values of the Yukawa couplings $Y$:
\[
Y_u=3.7\cdot 10^{-17} {\rm TeV}^{-3},\ \ Y_d=2.1\cdot 10^{-20}{\rm
TeV}^{-3}, \ \ Y_l=1.8\cdot 10^{-21}{\rm TeV}^{-3}.
\]

It is  worth noting that we do not obtain any additional hierarchy: all
dimensionless combinations of the our parameters are of order unity. For
instance, $g_u^{1/5}/\sqrt{\lambda }=0.4$,  $Y_u^{1/3}/g_u^{1/5}=1.3$,
etc.

\section{Conclusions}
In a class of multidimensional models with one vector-like fermionic
family, the low-energy effective theory describes three chiral
families in four dimensions. Hierarchy of fermionic masses appears due to
different profiles of the fermionic wave functions in  extra
dimensions.  In this paper we have investigated numerically one of the
models of this kind. Namely, we considered the model suggested in
~\cite{LT} and ~\cite{LTF}. We have shown that hierarchical mass pattern
indeed appears in the model ~\cite{LTF} with the gauged $U_g(1)$ group
which is responsible for the forming of Abrikosov--Nielsen--Olesen vortex,
but the dependence of the mass matrix on the parameters is slightly
different than it was presented in the Refs.~\cite{LT,LTF}. On the other
hand, in the case of global $U_g(1)$ unnatural  mass pattern appears.
In the gauge case we found the set of the parameters of the model which
reproduces well all known fermion masses and mixings, without hierarchy
among parameters.  Moreover, in SM quark sector we have reproduced the
nine known values (six masses and three mixing angles)  by the variation
of the seven parameters of the model: three couplings $g_{u,d,q}$, two
$\epsilon _{u,d}$, and two Yukawa couplings $Y_{u,d}$.

\acknowledgments
We are indebted to V.Rubakov, S.Troitsky, and J.-M. Frere for numerous
helpful discussions. We thank F.Bezrukov and A.Kuznetsov for the help in
numerical simulations.  This work is supported in part by RFFI grants
99-02-18410a, 01-02-06033, by CRDF award RP1-2103, by the Council of
Presidential Grants and State Support of Leading Scientific Schools, grant
00-15-96626, and by the programme SCOPES of the Swiss National Science
Foundation, project No. 7SUPJ062239, financed by Federal Department of
Foreign affairs.

\end{document}